\documentstyle[twocolumn,prl,aps]{revtex}            

\input psfig

\begin{document}
\draft

\twocolumn[\hsize\textwidth\columnwidth\hsize\csname 
@twocolumnfalse\endcsname                            


\title{Non-Arrhenius Behavior of Surface Diffusion 
Near a Phase Transition Boundary}

\author{I. Vattulainen,$^{1,2}$
        J. Merikoski,$^{1,2,3}$
        T. Ala-Nissila,$^{1,2,4}$ and 
        S. C. Ying$^{2}$}

\address{
$^1$Helsinki Institute of Physics,  
        P.O. Box 9 (Siltavuorenpenger 20 C), FIN--00014  
        University of Helsinki, Finland \\
$^2$Department of Physics, Box 1843, Brown University, 
        Providence, R.I. 02912, U.S.A. \\
$^3$Department of Physics, University of Jyv\"askyl\"a,  
        P.O. Box 35, FIN--40351 Jyv\"askyl\"a, Finland \\
$^4$Laboratory of Physics,  
        Tampere University of Technology, P.O. Box 692,  
        FIN--33101 Tampere, Finland
}

\date{\today}

\maketitle

\begin{abstract}
We study the non-Arrhenius behavior of surface diffusion near the second-order
phase transition boundary of an adsorbate layer. In contrast to expectations 
based on macroscopic thermodynamic effects, we show that this behavior can be 
related to the average microscopic jump rate which in turn is determined by 
the waiting-time distribution $W(t)$ of single-particle jumps at short times. 
At long times, $W(t)$ yields a barrier that corresponds to the rate-limiting 
step in diffusion. The microscopic information in $W(t)$ should be accessible 
by STM measurements. 
\end{abstract}


\vskip2pc]                                       


The migration of atoms and molecules is one of the most important  
processes taking place on solid surfaces. It appears in many phenomena  
such as catalytic reactions and surface growth that are important  
for practical applications \cite{applic}. In most experimental and  
theoretical studies of the surface diffusion constant $D$, its  
temperature dependence is analyzed through an assumed Arrhenius form,  
where $D$ is written as a product of an entropic prefactor $D_0$ and 
a term $\exp(-E_A^D/k_B T)$ describing thermally activated jumps over 
an energy barrier $E_A^D$. Although the Arrhenius form can be derived 
from microscopic considerations in some special cases \cite{Ala92,Che94}, 
a rigorous justification for its use in interacting systems at finite 
coverages is not available. Further, even in the cases where $D$ 
appears to have an Arrhenius temperature dependence over a finite 
temperature range, its microscopic interpretation may not always be 
clear. This is because for an interacting system, there may be many
microscopic activation barriers. Thus the value of the measured 
effective diffusion barrier $E_A^D$ must result from some complex 
average of all of them, and does not refer to any microscopic process 
in particular \cite{Gom90}.

In fact, the values for $D_0$ and $E_A^D$ can be strongly 
temperature-dependent indicating a region of non-Arrhenius behavior. 
This becomes especially pronounced near surface phase transition 
boundaries, where rapid variations of $D$ have been observed in 
experiments \cite{Gom90,Nau85,Zhd91} and computer simulations 
\cite{Ala92,Sad83}. Such rapid changes are often accompanied by the 
well-known ``compensation'' effect \cite{Mey37}, where an apparent 
increase in $E_A^D$ is compensated by an increase in the prefactor $D_0$ 
\cite{Zhd91}. However, in most cases the underlying reasons for 
non-Arrhenius behavior are not understood. It is the purpose of the present 
work to study these issues near a second-order phase transition 
in a surface adsorbate layer. We show that in contrast to the common 
folklore that an anomalous temperature dependence in $D$ near $T_c$ 
would be predominantly due to non-local thermodynamic effects, it can 
be explained by the microscopic single-particle jump rate $\Gamma$. 
This quantity is determined by the short-time behavior of the 
waiting-time distribution $W(t)$ for single-particle jumps. Moreover, 
we show that for long times, $W(t)$ yields an effective activation 
barrier that corresponds to the rate-limiting step in diffusion. Thus 
$W(t)$ provides a connection between microscopic and macroscopic aspects 
of diffusion. Further, it is experimentally available through {\it e.g.} 
STM measurements \cite{Swa96}.

In this Letter, we have carried out Monte Carlo (MC) simulations  
for a model of oxygen on the W(110) surface \cite{Vat96,Sah88}. In this 
system, the substrate remains unreconstructed \cite{Bru90}, the 
oxygen atoms have well-defined adsorption sites \cite{Elb97}, and 
desorption of oxygen occurs only at temperatures 1600 K or above 
\cite{Bru90}. Therefore, this system is very suitable for simulation 
studies using a lattice-gas description. We use the 
lattice-gas model constructed by Sahu {\em et al.} \cite{Sah88} to 
describe the main features of the phase diagram. The Hamiltonian 
includes pair interactions up to fifth nearest neighbors and some 
three-body interactions \cite{Sah88}, the attractive ones being 
the dominant. We concentrate on results for the coverage 
$\theta = 0.45$ over a wide temperature range. For this coverage at a 
low temperature, the adlayer is in the ordered p($2\times 1$) phase,  
while at $T_c \approx$ 710 K it undergoes a second-order transition 
\cite{Sah88} to a disordered phase \cite{Wan78}. For details of the 
model and MC simulations, see Refs.~\cite{Vat96,Vat96b}.

\begin{figure}[htb]
\null
\hspace{-2.0cm}\psfig{file=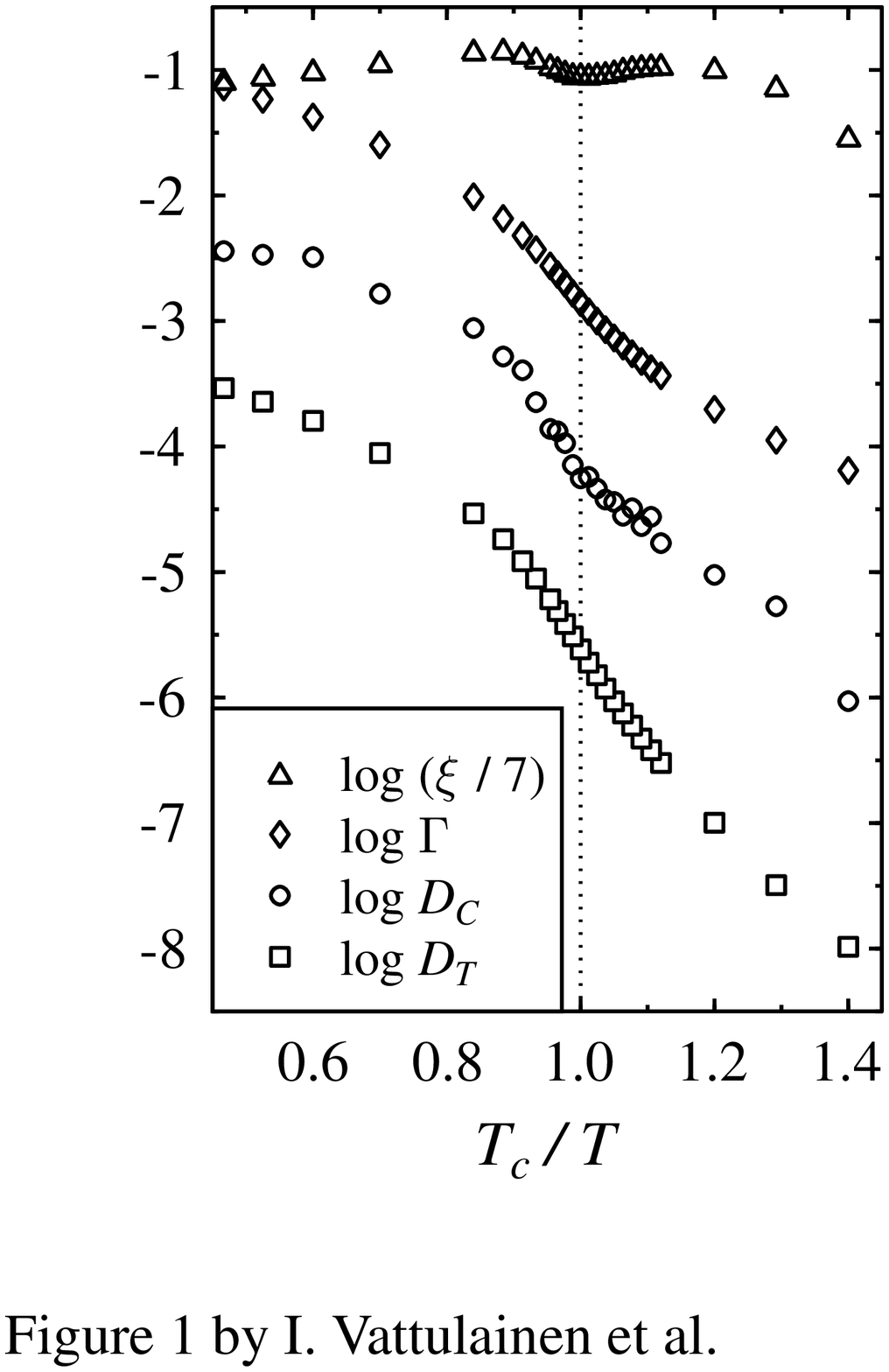,width=9.0cm,bbllx=-1pt,bblly=180pt,bburx=512pt,bbury=720pt,clip=}
\caption{
Results for $D_T$ and $D_C$ as a typical Arrhenius plot in the O/W(110) 
system at $\theta = 0.45$. 
We also show the behavior of the average transition rate $\Gamma$ 
and the thermodynamic factor $\xi$. 
The quantities $D_T$, $D_C$, and $\Gamma$ have been made dimensionless 
by expressing them in units of $a^2/t_0$, $a^2/t_0$, and $1/t_0$, 
respectively, where $a$ is the lattice constant 
and $t_0$ is one Monte Carlo step per particle.  
The value of $\xi$, which is a dimensionless quantity, 
has been scaled by a factor of seven to make the graphic 
representation more readable while other quantities are 
expressed directly in the units given above.
The critical temperature of the order-disorder phase transition is 
denoted by $T_c$ and a dotted line.\label{figure1}}
\end{figure}

Our simulation results for the tracer and collective diffusion 
coefficients $D_T$ and $D_C$ (for definitions see {\it e.g.} 
Ref.~\cite{Gom90}), respectively, are given in Fig.~\ref{figure1}.  
We first note that their qualitative behavior is similar and that the  
effective diffusion barrier $E_A^D$ defined as  
\begin{equation} 
\label{Eq:EaArrh} 
E_A^D \equiv - \frac{\partial (\log D)}{\partial (1/k_B T)}  
\end{equation} 
is approximately constant at low and high temperatures away from $T_c$.   
This implies that the diffusion constants obey simple Arrhenius 
behavior. Near $T_c$, however, the temperature dependence of the 
diffusion constants is strongly non-Arrhenius.

In the Green-Kubo formalism \cite{Gom90}, the expression for $D_C$ 
contains a thermodynamic factor $\xi$ inversely proportional to the 
compressibility, which is governed by the {\em global} number 
fluctuations of the adlayer. It is often assumed that a possible 
non-Arrhenius behavior of $D_C$ near the phase transition boundary is 
predominantly due to the critical behaviour of this factor \cite{Sad83}. 
We show below that this is not the case here: The non-Arrhenius behavior 
of both $D_T$ and $D_C$ has a dynamic origin and can be traced back 
to the temperature dependence of the {\em local} single-particle jump 
rate $\Gamma$. To demonstrate this, we show in Fig.~\ref{figure1} the 
temperature dependence of $\xi$ and $\Gamma$ as well as that of $D_T$ 
and $D_C$. It can be seen from Fig.~\ref{figure1} that while the 
diffusion constants have a turning point and sharp temperature 
variations close to $T_c$, the thermodynamic factor $\xi$ has only a 
relatively weak temperature dependence in this region and cannot account 
for the non-Arrhenius behavior of the diffusion constants. On the other 
hand, the single-particle jump rate $\Gamma$ has exactly the same 
behavior near $T_c$ as $D_T$ and $D_C$. These observations can be 
understood theoretically within the dynamical mean field theory (DMF) 
\cite{Hje97}, which yields $D_T \propto \Gamma$ and 
$D_C \propto \xi \Gamma$. We can conclude that the strong temperature 
dependence of both $D_T$ and $D_C$ near $T_c$ is indeed of the same 
dynamic origin, coming from the average single-particle transition 
rate $\Gamma$.

\begin{figure}
\hspace{-2.0cm}\psfig{file=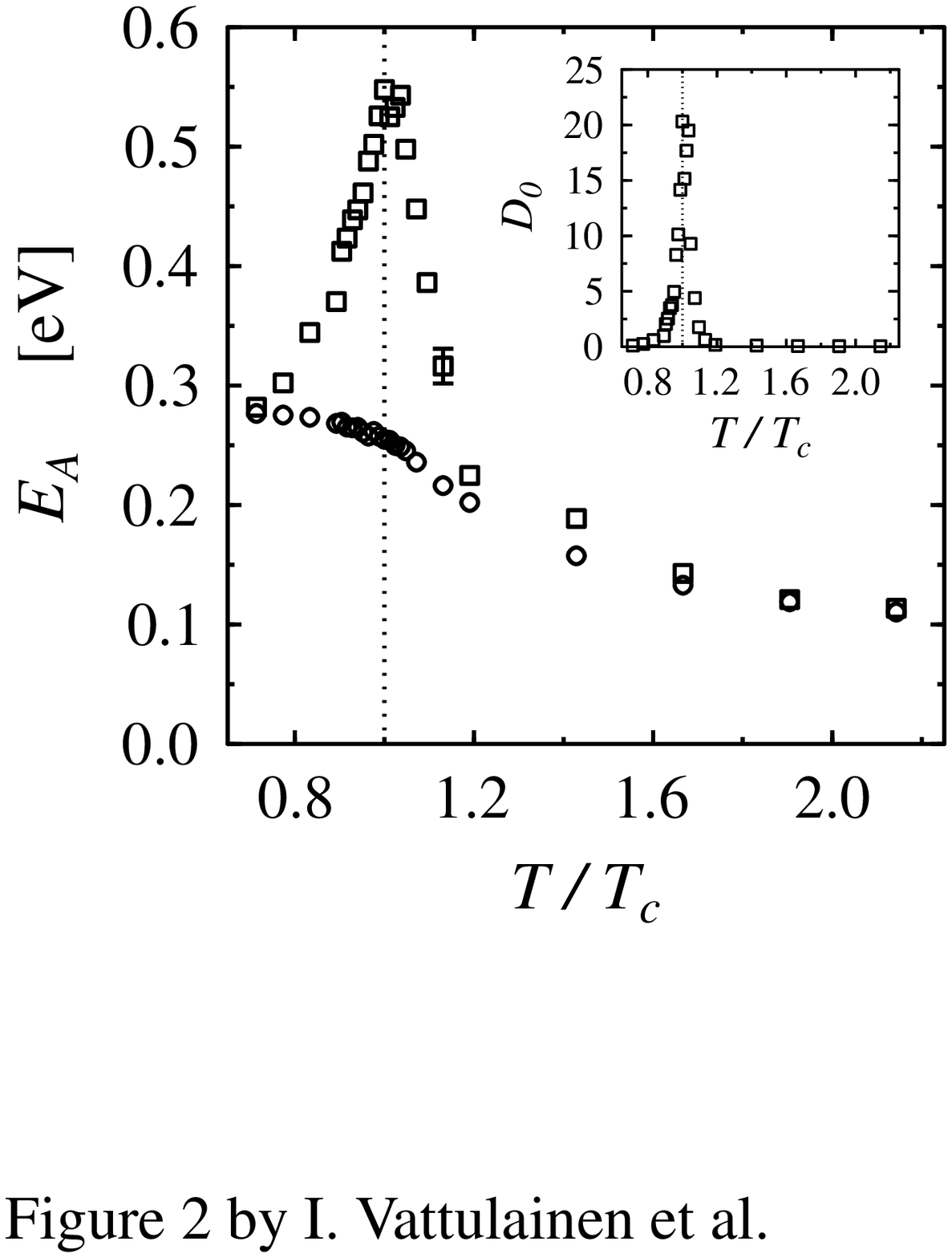,width=9.0cm,bbllx=-1pt,bblly=220pt,bburx=512pt,bbury=635pt,clip=}
\caption{
Results for the effective activation barriers. The squares denote results  
based on the Arrhenius form (see Eq.~(\protect\ref{Eq:EaArrh})) for tracer  
diffusion, while open circles represent the data based on the tail of 
$W(n)$. Behavior of the prefactor $D_0$ is illustrated in the inset. 
The critical temperature is denoted by a dotted line.\label{figure2}}
\end{figure}

We next focus on the effective diffusion barrier $E_A^D$ as extracted 
from Eq.~(\ref{Eq:EaArrh}) for $D_T$. As shown by squares in 
Fig.~\ref{figure2}, $E_A^D$ has a sharp peak near $T_c$. This peak in 
$E_A^D$ is accompanied by a strong increase in the value of the 
corresponding prefactor $D_0$ shown in the inset of Fig.~\ref{figure2}. 
This is yet another example of the well-known compensation effect 
\cite{Gom90,Mey37}. Here the compensation simply results from the fact 
that when the temperature dependence is non-Arrhenius, there is no 
unique way of separating the prefactor and the barrier contributions.  
Since the temperature dependence of the diffusion constant itself 
near $T_c$ is smooth and nonsingular, any dramatic change in the 
temperature dependence of the effective barrier $E_A^D$ {\it must} be 
followed by a corresponding change in the effective prefactor $D_0$. We 
note that the same phenomenon occurs for collective diffusion as well.

To understand the observed strong temperature variation of $E_A^D$ 
near $T_c$, we need to consider the energetics of the microscopic 
jump processes which determine the average jump rate $\Gamma$. At finite 
coverages, there is a very complex distribution $P(E_a)$ for the 
instantaneous activation barriers $E_a$ \cite{E_a} which an adatom needs 
to overcome in a jump attempt from one configuration to another. At 
high $T$, $P(E_a)$ is strongly peaked at small values of $E_a$, while at 
low temperatures the situation is completely the opposite \cite{Vat97}. 
The change in the distribution takes place around $T_c$, thus characterizing  
the ordering of the adlayer as the temperature is decreased below $T_c$. 
This change in turn results in a strong temperature dependence of the 
average transition rate $\Gamma$ around $T_c$, as shown in 
Fig.~\ref{figure1}. We point out that the instantaneous activation 
barriers $E_a$ cannot explain the peak of the effective barrier $E_A^D$ 
in Fig.~\ref{figure2}, since the largest value of $E_a$ in our model 
system is only about 0.4 eV \cite{Vat97}. Thus, the peak does not 
refer to any microscopic rate-limiting process. Instead, it arises 
from an entropic contribution \cite{TST} to $\Gamma$ which has a strong 
temperature dependence in the vicinity of $T_c$.


To gain more insight into the microscopic dynamical processes and the 
anomalous temperature dependence near $T_c$, we next introduce the 
waiting-time distribution $W(t)$ of single-particle jumps \cite{Hau87}. 
Suppose a single particle (in the presence of other particles) had 
performed its last transition at time $t = 0$. Then $W(t)$ is the 
probability density that the particle in question performs its next 
transition at time $t$ after it remained still until $t$. Here the most 
practical definition of ``time'' in the MC simulations is to consider 
the time scale as the number of jump attempts of the particle, denoted 
by $n$. Then the waiting-time distribution is simply $W(n)$. This 
provides a direct connection with the dynamic jump rate $\Gamma$ discussed 
above via
\begin{equation} 
\label{Eq:Gamma} 
\langle n \rangle \equiv \frac{1}{\Gamma} =  \sum_{n=1}^{\infty} n\, W (n), 
\end{equation} 
where $\langle n \rangle$ is the average waiting-time of the particle.

\begin{figure}
\hspace{-2.0cm}\psfig{file=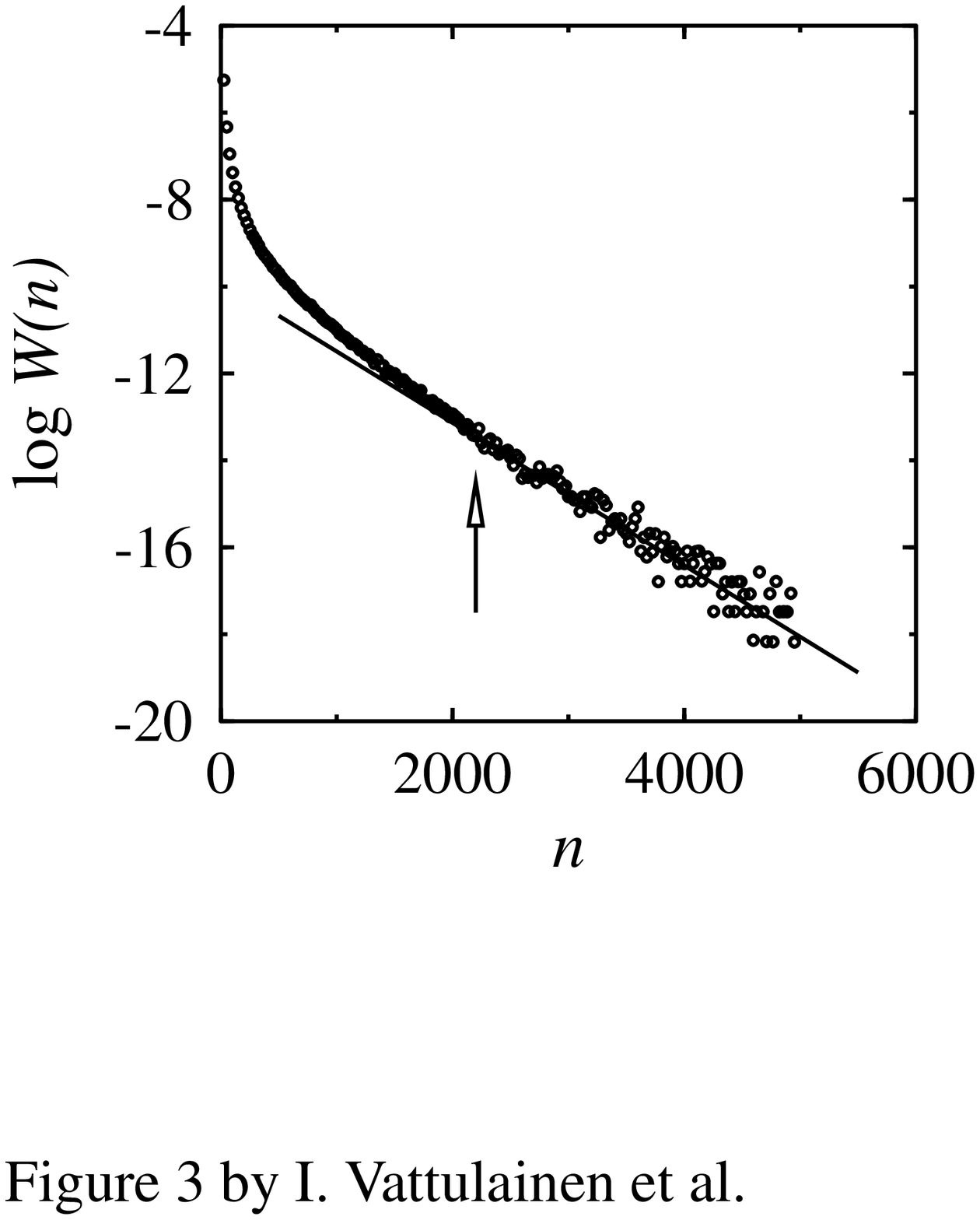,width=9.0cm,bbllx=-1pt,bblly=220pt,bburx=522pt,bbury=635pt,clip=}
\caption{
An example of a waiting-time distribution $W(n)$ at a temperature of  
0.774 $T_c$ showing an exponential decay at long times. For clarity,  
only some of the data points are shown here. The full curve is an 
exponential fit to the tail of $W(n)$. The approximate crossover time 
$n_{\rm co}$ for the crossover from the small-time regime to the 
asymptotic long-time regime \protect\cite{Crossover} is indicated 
by an arrow.\label{figure3}}
\end{figure}

At very long times, we expect $W(t)$ to decay as $W(t) \sim \exp(-t/\tau)$. 
Here the characteristic time $\tau$ describes the longest time scale 
among the various microscopic processes, which constitutes the 
rate-limiting  factor for mass transport. This expected exponential 
decay at long times  is indeed observed for our model system, as 
demonstrated in Fig.~\ref{figure3}. We can then define an effective 
activation barrier $E_A^W$ via $\tau$ by considering the jump probability 
$p = 1/\tau = p_0 \exp(- E_A^W / k_B T)$. As shown by circles in 
Fig.~\ref{figure2}, the activation barrier $E_A^W$ extracted from the 
asymptotic region of $W(t)$ decreases monotonically with increasing 
temperature, and agrees with the effective diffusion barrier $E_A^D$ 
extracted from an Arrhenius analysis of $D_T$ far from $T_c$. Additional 
studies in our model system \cite{Vat97} indicate that the value of $E_A^W$ 
is closely related to the instantaneous activation barrier characterizing 
the dominant microscopic processes. In our model the microscopic barriers 
have a maximum value of about 0.4 eV and thus the barrier $E_A^W$ does 
not have the sharp peak displayed by the effective diffusion barrier $E_A^D$.

\begin{figure}
\hspace{-2.0cm}\psfig{file=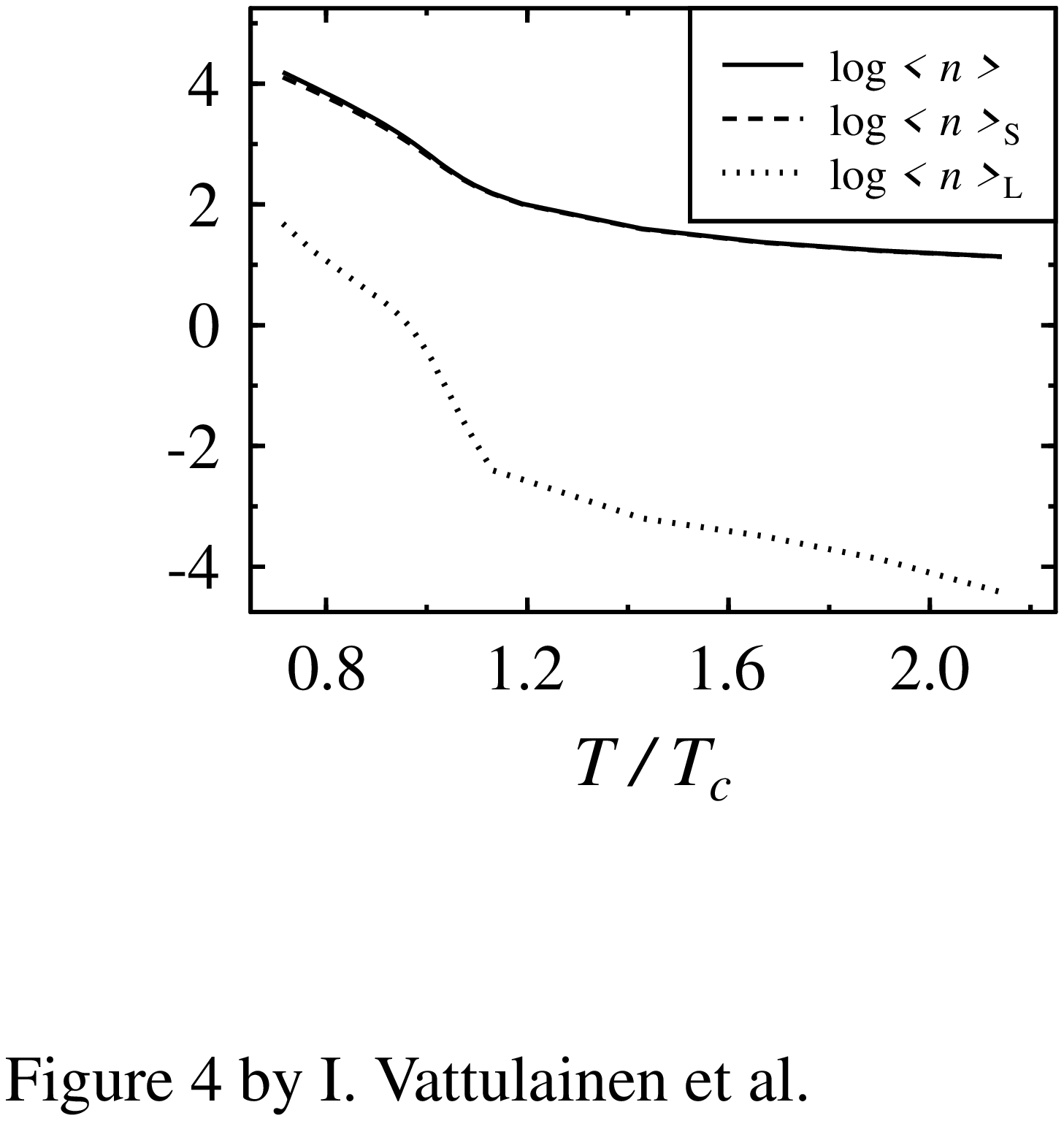,width=9.0cm,bbllx=-1pt,bblly=215pt,bburx=542pt,bbury=575pt,clip=}
\caption{
Comparison of the short-time contribution $ \langle n \rangle_{\rm S} $
and the long-time contribution $ \langle n \rangle_{\rm L} $
to the average waiting-time $ \langle n \rangle $. 
The slight increase of $ \langle n \rangle_{\rm L} $ at small $T$ is 
due to $n_{\rm co}$ whose value is difficult to determine accurately 
at very low temperatures.
The quantities $ \langle n \rangle_{\rm S} $, $ \langle n \rangle_{\rm L} $,
and $ \langle n \rangle $ are all expressed in units of one Monte Carlo
step per particle.
\label{figure4}}
\end{figure}

It turns out that the temperature dependence of the barrier $E_A^D$ 
results mainly from the short-time behavior of $W(n)$. 
This is demonstrated by dividing the sum in Eq.~(\ref{Eq:Gamma}) 
into two parts, the first of which is the short-time contribution 
$\langle n \rangle_{\rm S} = \sum_{n=1}^{n_{\rm co}} nW(n)$. This 
quantity accounts for the contribution up to a crossover time 
$n_{\rm co}$, which separates the short-time regime from the asymptotic 
exponential decay. What remains is the long-time contribution 
$\langle n \rangle_{\rm L} = \langle n \rangle -\langle n \rangle_{\rm S}$. 
As expected, from Fig.~\ref{figure4} we observe that the short-time 
regime gives the dominant contribution to $\Gamma$. Further, the 
short-time regime of $W(n)$ is strongly affected by the critical 
fluctuations, being mainly responsible for the anomalous temperature 
dependence of the diffusion constants near $T_c$.


To summarize, within the present model of O/W(110), the non-Arrhenius 
behavior near $T_c$ was found to have a mainly dynamic origin, reflecting 
the dependence of the single-particle jump rate $\Gamma$ on the  
critical fluctuations close to $T_c$. Surprisingly, in our studies the 
thermodynamic factor gives only a minor contribution to the temperature 
dependence of the collective diffusion constant $D_C$, and the anomalous 
temperature dependence for both $D_T$ and $D_C$ results from the dynamic 
factor $\Gamma$. We find that the single-particle waiting-time 
distribution $W(t)$ gives the most detailed picture of the microscopic 
processes. 
It has been recently demonstrated by Swartzentruber \cite{Swa96} that this
distribution function in the presence of several different microscopic
activation barriers can indeed be measured using the STM. From the 
long-time tail of $W(t)$ one can obtain information on the 
energetics of the rate-limiting processes of diffusion in the form of an 
effective activation barrier $E_A^W$. On the other hand, the temperature 
variation of the effective diffusion barrier $E_A^D$ directly reflects 
that of the microscopic jump rate $\Gamma$, and depends not only on the 
long-time tail of $W(t)$ but also on its short-time behavior. In the 
short-time regime near $T_c$, $W(t)$ is strongly 
affected by the critical fluctuations. 
The fluctuations there lead to a strong temperature dependence of 
the transition entropy and an additional contribution to the 
effective barrier $E_A^D$.


I. V. thanks the Neste Co. Foundation and the Jenny and Antti Wihuri  
Foundation. J. M. is supported by the Finnish Cultural Foundation. 
This research has also been partially supported by a grant from the 
office of Naval Research (S. C. Y. and J. M.) and by the Academy of 
Finland (J. M. and T. A.-N.). Finally, computing resources of the 
University of Helsinki are gratefully acknowledged.

\begin{references}
\bibitem{Crossover}  
In this work, the crossover time $n_{\rm co}$ was chosen to be the 
value of $n$, at which the relative deviation of $\log W(n)$ from 
its asymptotic form was half percent \protect\cite{Vat97}.
\end{references}
\end{document}